# THE MEANING OF THE FINE STRUCTURE CONSTANT

Robert L. Oldershaw

Amherst College

Amherst, MA 01002  USA

rloldershaw@amherst.edu

**Abstract:**  A possible explanation is offered for the longstanding mystery surrounding the meaning of the fine structure constant.  The reasoning is based on a discrete self-similar cosmological paradigm that has shown promise in explaining the general scaling properties of nature's global hierarchy.  The discrete scale invariance of the paradigm implies that "strong gravity" governs gravitational interactions *within* atomic scale systems.  A new atomic scale gravitational coupling constant ($G_{-1}$) is derived with a value of $\approx 10^{38}\ G_0$, where $G_0$ is the conventional Newtonian gravitational constant. $G_{-1}$ is then used to calculate a revised Planck scale.  Given $G_{-1}$ and the revised Planck mass, one can demonstrate that within the context of the discrete self-similar paradigm the fine structure constant is the ratio of the strengths of the unit electromagnetic interaction and the unit gravitational interaction *within* atomic scale systems.

1. **Introduction**

Arnold Sommerfeld introduced the fine structure constant into physics in the 1920s in order to account for the relativistic splitting of atomic spectral lines. Following its introduction, the fine structure constant ($\alpha$) was discovered to be a fairly ubiquitous constant, occurring frequently in the physics of atomic scale systems and quantum electrodynamics. One of the strangest things about $\alpha$, given its importance and ubiquity, is the fact that for over 80 years it has remained enigmatic. The brilliant physicist Wolfgang Pauli famously quipped: 'When I die, my first question to the devil will be: What is the meaning of the fine structure constant?' Recently, eight decades after its introduction into physics, analyses based on a discrete fractal model of the cosmos have yielded a unique and natural explanation for the origin of $\alpha$.

The arguments presented below are based on the Self-Similar Cosmological Paradigm (SSCP)[1-6] which has been developed over a period of more than 30 years, and can be unambiguously tested via its definitive predictions [1,4] concerning the nature of the galactic dark matter. Briefly, the discrete self-similar paradigm focuses on nature's fundamental organizational principles and symmetries, emphasizing nature's *intrinsic hierarchical organization* of systems from the smallest observable subatomic particles to the largest observable superclusters of galaxies. The new discrete fractal paradigm also highlights the fact that nature's global hierarchy is *highly stratified*. While the observable portion of the entire hierarchy encompasses nearly 80 orders of magnitude in mass, three relatively narrow mass ranges, each extending for only about 5 orders of magnitude, account for ≥ 99% of all mass observed in the cosmos. These dominant mass ranges: roughly $10^{-27}$ g to $10^{-22}$ g, $10^{28}$ g to $10^{33}$ g and $10^{38}$ g to $10^{43}$ g, are referred to as the *Atomic, Stellar and Galactic Scales*, respectively. The cosmological Scales constitute the discrete self-similar scaffolding of the observable portion of nature's quasi-continuous hierarchy. At present the number of Scales cannot be known, but for reasons of natural philosophy it is tentatively proposed that there are a denumerably infinite number of cosmological Scales, ordered in terms of their intrinsic ranges of space, time and mass scales. A third general principle of the new paradigm is that the *cosmological Scales are rigorously self-similar to one another*, such that for each class of fundamental particles, composite systems or physical phenomena on a given Scale there is a corresponding class of particles, systems or phenomena on all other cosmological Scales. Specific self-similar analogues from different Scales have rigorously analogous morphologies, kinematics and dynamics. When the general self-similarity among the discrete Scales is *exact*, the paradigm is referred to as Discrete Scale Relativity [5] and nature's global space-time geometry manifests a new universal symmetry principle: *discrete scale invariance*.

Based upon decades of studying the scaling relationships among analogue systems from the Atomic, Stellar and Galactic Scales,[1-6] a close approximation to nature's self-similar Scale transformation equations for the length (L), time (T) and mass (M) parameters of analogue systems on neighboring cosmological Scales $\Psi$ and $\Psi$-1, *as well as for all dimensional constants*, are as follows.

$$L_\Psi = \Lambda L_{\Psi-1} \qquad (1)$$

$$T_\Psi = \Lambda T_{\Psi-1} \qquad (2)$$

$$M_\Psi = \Lambda^D M_{\Psi-1} \qquad (3)$$

The self-similar scaling constants $\Lambda$ and D have been determined empirically and are equal to $\cong 5.2 \times 10^{17}$ and $\cong 3.174$, respectively.[2,3] The value of $\Lambda^D$ is $1.70 \times 10^{56}$. Different cosmological Scales are designated by the discrete index $\Psi$ ($\equiv$ …, -2, -1, 0, 1, 2, …) and the Atomic, Stellar and Galactic Scales are usually assigned $\Psi = -1$, $\Psi = 0$ and $\Psi = +1$, respectively.

The fundamental self-similarity of the SSCP and the recursive character of the discrete scaling equations suggest that nature is an infinite discrete fractal, in terms of its morphology, kinematics and dynamics. The underlying principle of the paradigm is discrete scale invariance and the physical embodiment of that principle is the discrete self-similarity of nature's physical systems. Perhaps the single most thorough and accessible resource for exploring the SSCP is the author's website.[6]

## 2. "Strong Gravity"

Because the discrete self-similar scaling of the new paradigm applies to *all dimensional parameters*, the Scale transformation equations also apply to dimensional "constants." It has been shown[5] that the gravitational coupling constant $G_\Psi$ scales as follows.

$$G_\Psi = [\Lambda^{1-D}]^\Psi G_0 \quad , \qquad (4)$$

where $G_0$ is the conventional Newtonian gravitational constant. Eq. (4) results from the $L^3/MT^2$ dimensionality of $G_\Psi$ and the self-similar scaling rules embodied in Eqs. (1) - (3). Therefore the Atomic Scale value $G_{-1}$ is $\Lambda^{2.174}$ times $G_0$ and equals $\cong 2.18 \times 10^{31} cm^3/g\ sec^2$.

The value of the gravitational coupling constant has been tested on a variety of size scales, but it has *never* been empirically measured *within* an Atomic Scale system. To be perfectly clear on this point, the distinction between the appropriateness of using $G_0$ or $G_{-1}$ as the correct gravitational coupling constant is less determined by size scales than by whether the region of interest is within an Atomic Scale system, or exterior to Atomic Scale systems. The possibility that the Atomic Scale gravitational coupling factor is on the order of $10^{38}$ times larger than its counterpart within a Stellar Scale system has recently found support in successful retrodictions of the proton mass and radius using the geometrodynamic form of Kerr-Newman solutions to the Einstein-Maxwell equations.[7]

### 3. A Revised Planck Mass

The conventional Planck scale is based on the use of $G_0$ to determine the numerical values of the Planck mass, length and time. If the SSCP is the correct paradigm for developing a more unified physics, then a revised Planck scale based on $G_{-1}$ yields the following values.

$$\text{Planck length} = [\hbar G_{-1}/c^3]^{1/2} = 2.93 \times 10^{-14} \text{ cm} \approx 0.4 \text{ proton radius} \quad (5)$$

$$\text{Planck mass} = [\hbar c/G_{-1}]^{1/2} = 1.20 \times 10^{-24} \text{ g} \approx 0.7 \text{ proton mass} \quad (6)$$

$$\text{Planck time} = [\hbar G_{-1}/c^5]^{1/2} = 9.81 \times 10^{-25} \text{ sec} \approx 0.4 \text{ (proton radius/c)} \quad (7)$$

### 4. The Fine Structure Constant

The conventional definition of the fine structure constant ($\alpha$) is:

$$\alpha = e^2 / 4\pi\varepsilon_0 \hbar c = 7.297 \times 10^{-3}, \quad (8)$$

where e is the unit electromagnetic charge, $\varepsilon_0$ is the permittivity constant, $\hbar$ is Planck's constant divided by $2\pi$, and c is the velocity of light. Since $\alpha$ is *dimensionless*, it seems natural to expect that it is the ratio of two quantities with the same dimensionality, as in the case of $\pi$. It also seems very reasonable to group the constants of Eq. (8) in the following manner.

$$\alpha = [e^2/4\pi\varepsilon_0] / [\hbar c] \qquad (9)$$

From the derivation of the *revised* Planck mass, $M_{pl}$, we know that:

$$M_{pl} = [\hbar c / G_{-1}]^{1/2} = 1.20 \times 10^{-24} \text{ g}$$

and that

$$\hbar c = G_{-1} (M_{pl})^2 . \qquad (10)$$

Eq. (10) can be verified numerically and holds good to a reasonable level of significance, given the small unavoidable uncertainty in determining $\Lambda^D$ empirically. When we substitute $G_{-1} (M_{pl})^2$ for $\hbar c$ in Eq. (9), we find that:

$$\alpha = [e^2 / 4 \pi \varepsilon_0] / [G_{-1} (M_{pl})^2] . \qquad (11)$$

Although one could in principle arbitrarily define any gravitational coupling factor $G_n$ and use it to derive a correlated "Planck mass" $M_n$, such that the $[G_n, M_n]$ pair would satisfy Eq. (11), the SSCP's identification of the specific $[G_{-1}, M_{pl}]$ pair is unique and based on an analysis of a large amount of empirical data for Atomic Scale and Stellar Scale systems.[1-6] Additional evidence for the uniqueness of the specific $[G_{-1}, M_{pl}]$ pair as the correct parameters for Eq. (11) will be presented in section 5 below.

## 5. The Meaning of α

The numerator of Eq. (11) is the square of the unit electromagnetic charge and the denominator is the square of the unit gravitational "charge" for Atomic Scale systems. Equivalently, the numerator can be interpreted as the strength of the unit electromagnetic interaction and the denominator can be interpreted as the strength of the unit gravitational interaction for Atomic Scale systems.

The uniqueness and appropriateness of the choice of the [$G_{-1}$, $M_{pl}$] pair for Eq. (11), as well as the proposed interpretation of α, are supported by the following results.

(a) As mentioned above, if $G_{-1}$ is the correct gravitational coupling factor *within* Atomic Scale systems, then the radius and mass of the proton obey the Kerr-Newman solution of the Einstein-Maxwell equations.[7]

(b) The SSCP's specific value of $M_{pl}$ is indistinguishable from the unique Kerr-Newman mass ($m^2 = a^2 + q^2$) that defines the boundary between solutions with or without event horizons.[8]

(c) When $G_{-1}$ is used as the gravitational coupling factor in the determination of the Gravitational Bohr Radius of the hydrogen atom, one gets the appropriate result of ~$2\pi$ times the conventional Bohr Radius.[9] In contrast, if one uses the Newtonian $G_0$, one gets a Gravitational Bohr Radius on the order of the size of the observable universe.

(d) If $G_{-1}$ is the appropriate gravitational coupling factor for Atomic Scale systems, then the alarming and enigmatic 120 orders of magnitude disparity between the vacuum energy densities of cosmology and high energy physics is reduced by at least 115 orders of magnitude, and may be removed entirely.[10]

Results (a) – (d) argue compellingly that the [$G_{-1}$, $M_{pl}$] pair and the proposed interpretation of $\alpha$ are unique and appropriate in the context of Atomic Scale systems. None of these potentially important results could have been achieved using $G_0$, or any other $G_n$ value, as the relevant gravitational coupling factor.

Without dying, or enlisting the aid of Pauli's devil, we have identified a natural and compelling answer to the longstanding mystery of the meaning of $\alpha$: *the fine structure constant is the ratio of the strengths of the fundamental unit electromagnetic and gravitational interactions.* Since $\alpha$ is dimensionless, the SSCP asserts that it has the same value on each of nature's discrete cosmological Scales.